\documentclass[prl,twocolumn,showpacs,superscriptaddress,floatfix]{revtex4}
\usepackage[dvips]{epsfig}
\usepackage{float}
\usepackage{amsfonts}
\usepackage{amsmath}
\usepackage{amssymb}
\usepackage{enumerate}
\begin{document}
\title{Terahertz surface plasmon polariton propagation and focusing on periodically corrugated metal wires}
\author{Stefan A. Maier}
\email{S.Maier@bath.ac.uk}
\author{Steve R. Andrews}
\affiliation{Centre for Photonics and Photonic Materials, Department of Physics, University of Bath, Bath BA2 7AY, UK}
\author{L. Mart\'{\i}n-Moreno}
\affiliation{Departamento de F\'{\i}sica de la Materia Condensada,
Universidad de Zaragoza-CSIC, E-50009 Zaragoza, Spain}
\author{F. J. Garc\'{\i}a-Vidal}
\email{fj.garcia@uam.es}
\affiliation{Departamento de F\'{\i}sica Te\'orica de la Materia
Condensada, Universidad Aut\'onoma de Madrid, E-28049 Madrid, Spain}
\date{\today}
\begin{abstract}
In this letter we show how the dispersion relation of surface plasmon polaritons (SPPs) propagating along a perfectly conducting wire can be tailored by corrugating its surface with a periodic array of radial grooves. In this way, highly localized SPPs can be sustained in the terahertz region of the electromagnetic spectrum. Importantly, the propagation characteristics of these spoof SPPs can be controlled by the surface geometry, opening the way to important applications such as energy concentration on cylindrical wires and superfocusing using conical structures.    
\end{abstract}
\maketitle
\setcounter{page}{1}
One of the major driving forces in the field of plasmonics is the ability to spatially confine electromagnetic energy at visible frequencies over distances significantly smaller than the wavelength. Surface plasmon polaritons (SPPs) and localized plasmons at metal/dielectric interfaces open up a previously inaccessible length scale for optical research, with promising applications in the miniaturization of photonic circuits \cite{Barnes,Ozbay}, near-field optics \cite{Girard} and single-molecule optical sensing \cite{Nie,Kneipp2}. 
It would be greatly advantageous to take concepts such as highly localized waveguiding \cite{Lamprecht,MaierNatMat, Bozhevolnyi} and superfocusing \cite{Stockman, Babadjanyan} to lower frequencies, particularly the THz regime \cite{Ferguson}, where plasmonics could enable near-field imaging \cite{Chen} and biosensing \cite{Nagel} with unprecedented sensitivity.
However, there is a basic problem: SPPs offer sub-wavelength field localization only for frequencies close to the intrinsic plasma frequency of the conductor, which for most metals is in the ultraviolet part of the spectrum. At THz frequencies on the other hand, metals resemble in many ways a perfect conductor, and the negligible penetration of the electromagnetic fields leads to highly delocalized SPPs akin to grazing-incidence light fields. In this frequency range, SPPs are also known as Sommerfeld-Zenneck waves \cite{Goubau}. Such weakly guided THz waves have recently been characterized on cylindrical metallic wires \cite{Wang,Wang2,Jeon}. While this waveguide geometry is in principle suitable for a variety of applications such as biological sensing in confined spaces \cite{Wang} or the local probing of materials \cite{Valk}, the radial extent of the fields over many wavelengths and the concomitant radiation loss due to bends, non-uniformities or nearby objects severely limit the efficiency and viability of such concepts.

In this letter we present a very promising route to achieving sub-wavelength confinement in metal wires at THz or microwave frequencies by showing that the dispersion relation of SPPs can be engineered at will by periodically structuring the cylindrical surface with grooves. In this way, the confinement does not rely on the finite conductivity of the wire, but is purely due to the surface structure. It is interesting to note that the viability of this approach was hinted at by Goubau in 1950 \cite{Goubau}. More recently, the idea of tailoring the topography of a {\it perfect} conductor to allow the existence of surface waves resembling the behavior of SPPs at optical frequencies was discussed in the context of two-dimensional hole lattices \cite{PendryScience} and one-dimensional groove arrays \cite{GarciaVidalSpoof} machined into flat interfaces. Experimental verification of these findings in the microwave regime has been recently reported \cite{Hibbins}. Due to their mimicking characteristics, these geometry-controlled surface waves were named  {\it spoof} SPPs.  Here we demonstrate that spoof SPPs can be sustained and exhibit a rich behavior in three dimensions, including the possibility of deep sub-wavelength energy concentration in cylindrical and superfocusing in conical structures. The length scales of the surface structures that we describe are within the limits of state-of-the-art laser-machining techniques that are used to make corrugated feedhorns for sub-millimeter radiation \cite{Lubecke}.

The inset of Fig. 1a shows a schematic picture of the envisioned structure. An array of radial grooves of depth $h=R-r$, width $a$ and lattice constant $d$ is machined into a perfectly conducting cylinder of radius $R$. Within the perfect conductor approximation, valid for metals in the microwave or THz ranges of the electromagnetic spectrum, the frequencies of the dispersion bands scale with the reciprocal size of the structure. Therefore, in Fig. 1 $d$ is used as the unit length. We are interested in calculating the dispersion relation (frequency, $\omega$, versus axial propagation constant, $k$) of TM-polarized waves propagating in the z-direction along the wire. First we apply a theoretical framework based on a modal expansion of the EM-fields. In order to yield quasi-analytical expressions for $\omega(k)$, we search for surface waves that have no azimuthal ($\phi$) variation, i.e. the azimuthal order $m$ is equal to zero, and when describing the EM-fields inside the subwavelength grooves, only the lowest eigenmode is considered in the expansion. In this way, in region II, $\bf{E}$ and $\bf{B}$ are zero everywhere except inside the radial grooves and $E_z$ can be written as:
 
\begin{equation}
E^{II}_z(\rho)=A J_0(g\rho)+B N_0(g\rho)  
\end{equation}

\noindent where $A$ and $B$ are constants, $J_0$ and $N_0$ are the zero-order Bessel and Neumann functions, respectively, and $g=\omega/c$, with $c$ being the velocity of light. Within this single mode approximation $E^{II}_{\rho}=0$. 

On the other hand, the z-component of the {\bf E}-field in region I (vacuum) can be expressed as:
\begin{equation}
E^{I}_z(\rho,z)=\sum_{n=-\infty}^{+\infty} C_n K_0(q_n \rho)e^{ik_n z}
\end{equation}
\noindent where $C_n$ are constants, $k_n=k+2\pi n/d$ takes into account diffraction effects and  $q_n=\sqrt{k_{n}^{2}-g^2}$. The radial dependence is controlled by $K_0$, the zero-order modified Neumann function, which displays a non-divergent behavior as $\rho \rightarrow \infty$. Note that, as we are interested in the non-radiative region of the dispersion relation, $k$ must be outside the light cone, i.e., $k > g$. The other non-zero components of $\bf{E}$ and $\bf{B}$ can be obtained straightforwardly from $E_z$.
The dispersion relation of the TM-polarized waves propagating along the cylinder is determined by imposing matching conditions on the parallel components of {\bf E} and {\bf B} at the interface between regions I and II and at the bottom of the grooves. Following this procedure yields a transcendental equation for $\omega(k)$:
\begin{equation}
\sum_{n=-\infty}^{+\infty}S_{n}^{2}\frac{g}{q_n}\frac{K_1(q_n R)}{K_0(q_n R)}\frac{N_0(g R)J_0(g r)-N_0(g r)J_0(g R)}{N_0(g r)J_1(g R)-N_1(g R)J_0(g r)}=1
\end{equation}
\noindent where $J_1$ and $N_1$ are the first-order Bessel and Neumann functions, respectively, and $K_1$ is the first-order modified Neumann function. $S_n$ is given by $S_n=\sqrt{a/d}\  \textrm{sinc} (k_{n}a/2)$.

\begin{figure}[t]
\begin{center}
\includegraphics[width=\columnwidth]{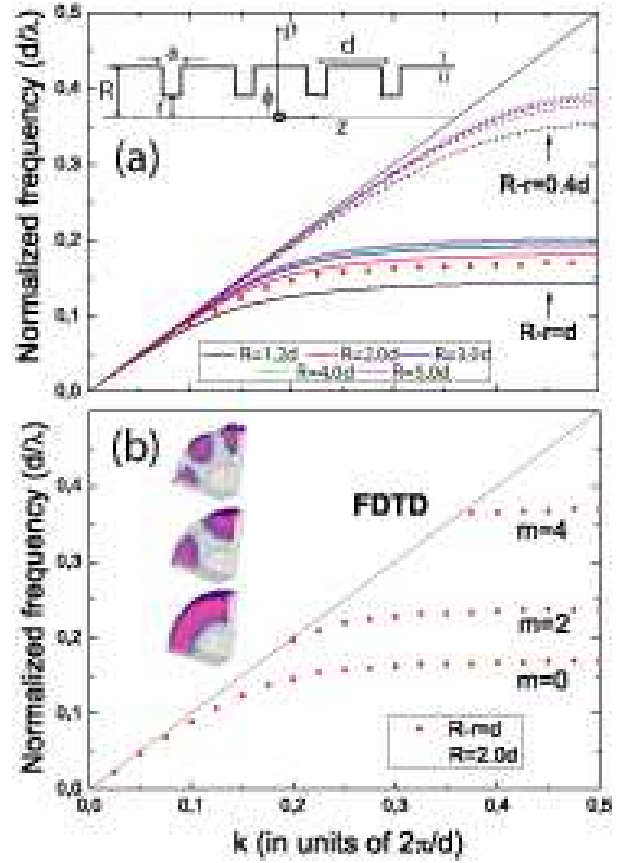}
\end{center}
\caption{(a) $\omega$ versus $k$ for the corrugated, azimuthally invariant wire geometry shown in the inset for two different cases: $h=R-r=0.4d$ (dashed lines) and $h=d$ (full lines). For each value of $h$, the dispersion relations calculated for increasing values of $R$ and $r$ ($h$ fixed) are presented. The dotted line shows the band obtained with a FDTD code for $h=d$ and $R=2d$. (b) Dispersion relation of the SPP bands obtained with the FDTD code for azimuthal modes $m=0$, $m=2$ and $m=4$. The insets show the amplitude of the E-field in a quarter cylinder for these three bands at $k=\pi/d$.}
\end{figure}

Figure 1a shows the dispersion relation calculated using Eq.(3) for SPPs propagating along a periodically corrugated, perfectly conducting cylinder. The inclusion of just $3$ or $4$ diffraction orders in the sum appearing in Eq.(3) yields sufficient convergence. The width of the grooves is $a=0.2d$ and two different depths of indentation are analyzed: $h=R-r=0.4d$ (dashed lines) and $h=d$ (full lines). For both depths, several values of the outer ($R$) and inner ($r$) radii are shown for fixed $h$. Figure 1a clearly shows that the inclusion of a periodic array of grooves on a perfectly conducting wire dramatically changes the dispersion relation of SPPs. Note that, within the perfect conductor approximation, SPP bands coincide with the light line for the non-corrugated geometry, and the modes are not bound. For the case of spoof SPPs, $\omega(k)$ departs significantly from the light line and this departure is greater when the indentations are deeper. This dispersion relation closely resembles the behavior of SPPs propagating along a metal wire at optical frequencies. In this last case, for $m=0$, $\omega(k)$ approaches $\omega_p /\sqrt{2}$ (where $\omega_p$ is the plasma frequency of the metal) at large $k$. For the spoof SPPs bands, the asymptotic frequency ($\omega_S$) is mainly controlled by the depth of the grooves. An analytical expression for $\omega_S$ can be obtained in the limit $R,r >>d$ and $\lambda \gg d,a$. By using the asymptotic expansions of the different Bessel and Neumann functions appearing in Eq.(3) and neglecting the diffraction orders, $k(\omega)$ becomes:

\begin{equation}
k=g \sqrt{1+\frac{a^2}{d^2} \tan^2  gh}   
\end{equation}

The asymptotic frequency $\omega_S=\pi c /2 h$ is inversely proportional to $h$, explaining the behavior of $\omega(k)$ observed in Fig. 1a when the depth of the radial grooves is varied.  
     
In order to check the accuracy of the approximations made in our quasi-analytical model, we have conducted finite-difference-time-domain (FDTD) numerical simulations on one unit cell of the wire structure. In Fig. 1a the spoof SPP band obtained with this code for the case of $h=d$ and $R=2d$ is also plotted (dotted line). The agreement between the band obtained from Eq.(3) and the FDTD result is remarkable. In Fig. 1b we present the $m=0,2,4$ bands (only for $k > g$) for the corrugated wire analyzed above calculated with the FDTD code. 
The spatial dependence of the amplitude of the {\bf E}-field for the three bands ($m=0,2,4$) evaluated at $k=\pi/d$, border of the first Brillouin Zone, is plotted in the inset. The highly localized character of the spoof SPPs along the radial direction is evident.

Once it has been shown how the bands of spoof SPPs can be engineered, we now focus on their light guiding properties in a wire tailored to operate at a frequency near 1 THz. In panel (a) of Fig. 2, we show the dispersion relation of the spoof SPPs bands obtained with our quasi-analytical approach for $a=10~\mu m$, $d=50~\mu m$, $r=50~\mu m$ and $R=100~\mu m$ (black line). For this set of geometrical parameters the asymptotic frequency, $f_S$, is equal to $1.1$ THz. The spatial variation of the {\bf E}-field associated to the spoof SPPs for different frequencies is displayed in panels (b-d) of Fig. 2. These pictures have been obtained by numerical simulations using the finite integration technique (FIT). The wire has been excited using an input port with a radially polarized broadband THz pulse, and multiple calculations with adaptive mesh refinement ensured well-converged results. For $f=0.6$~THz, the E-field is highly concentrated on the surface of the wire but this concentration is much stronger at a higher frequency, $f=1.0$~THz. This is due to the fact that in this last case the propagation constant is much longer than $\omega/c$ (see Fig.2a) resulting in a very large radial component of the wavevector, $q=\sqrt{k^2-g^2}$. For $f=1.2$~THz ($>f_S$), no spoof SPPs propagate along the corrugated wire and light is scattered at the input port. 

\begin{figure}[t]
\begin{center}
\includegraphics[width=\columnwidth]{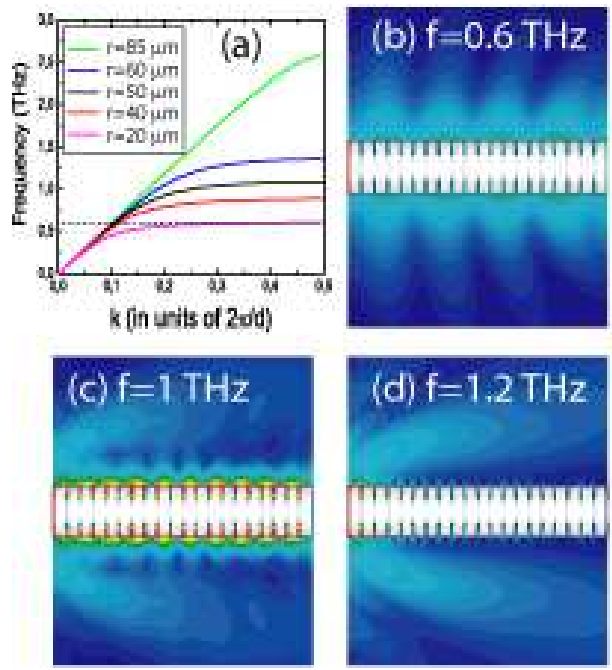}
\end{center}
\caption{(a) Dispersion relations of spoof SPPs for $a=10 \mu m$, $d=50 \mu m$ and $R=100 \mu m$ for five different values of the inner radii $r=85,60,50,40$ and $20 \mu m$. (b-d) Spatial variation of the {\bf E}-field amplitude (logarithmic scale) calculated using the finite integration technique on a wire with $a=10 \mu m$, $d=50 \mu m$, $R=100 \mu m$, and $r=50 \mu m$ for three different frequencies, $f=0.6,1.0$ and $1.2$ THz.}
\end{figure}

Both the high localization and low group velocity of spoof SPPs on corrugated metal wires are advantageous for applications in the routing of radiation with low bend losses and for sensing. The fact that the dispersion relation of the spoof SPPs can be easily tuned by engineering the geometry of the indentations opens up an even more fascinating possibility, the gradual energy concentration of radiation propagating along wires, and superfocusing of radiation at the tip of conical structures. 
 
Panel (a) of Fig. 2 gives us a clue on how THz radiation can be concentrated using periodically corrugated metal wires. Let us imagine a metal wire in which the outer radius is fixed ($R=100~\mu m$) and the inner one ($r$) is adiabatically (i.e., back-reflection and scattering are negligible) reduced, thus increasing the groove depth, from $r=100~\mu m$ at the beginning to $20~\mu m$ at the end of the wire. If the frequency of the surface wave is fixed (for example, $f=0.6~\text{THz}$, $\lambda=500~\mu m$, see dotted line in Fig.2a), it is expected that the group velocity of the mode will be gradually reduced as it propagates along the wire in the direction of decreasing $r$. Along with this change in the group velocity, the spatial extension of the mode in the radial direction would be severely reduced, as can be seen in Fig.3 (dotted lines), in which the dependence of the E-field concentration with $r$ is analyzed. Notice that these curves have been obtained with the quasi-analytical method, assuming that the wire is infinite and $r$ is fixed. To confirm the idea of gradual energy concentration, we have conducted FIT numerical simulations on a {\bf finite} straight wire of radius $R=100~\mu m$ and length $2$ mm containing $40$ grooves with $d=50~\mu m$. In this wire the inner radius $r$ decreases from $95~\mu m$ to $20~\mu m$ in steps of $5~\mu m$ every two lattice periods. This region is followed by a number of lattice periods with constant $r=20~\mu m$. The distribution of the E-field for $f=0.6$~THz is plotted on a logarithmic scale in the inset of Fig. 3. As expected from our previous discussion, the mode becomes increasingly confined to the surface with propagation distance, as shown in the main part of Fig. 3 (solid lines). At the end of the wire, the {\bf E}-field is concentrated on a region $10$ times smaller than the wavelength on the surface of the wire. The agreement between the quasi-analytical method and FIT validates the idea of concentrating light via an adiabatic reduction of $r$.  

\begin{figure}[t]
\begin{center}
\includegraphics[width=\columnwidth]{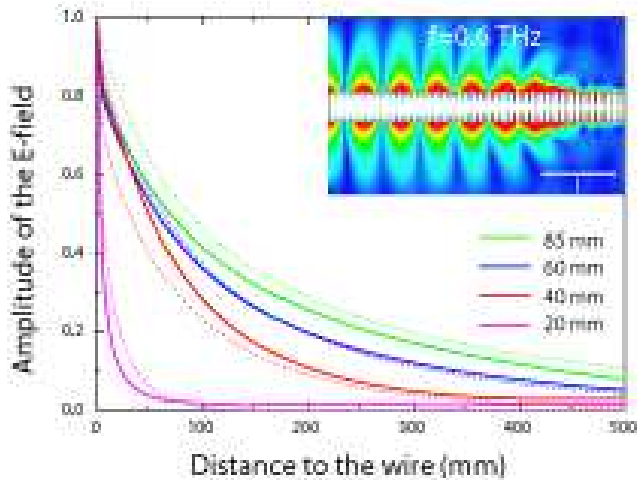}
\end{center}
\caption{Field concentration via adiabatic reduction of $r$. Distribution of the E-field, evaluated at $f=0.6$~THz,  along the radial direction at different locations of the wire corresponding to $r=85,60,40$ and $20 \mu m$ calculated using the quasi-analytical method for an infinite wire (dotted lines) and FIT for the real finite structure (solid lines) on a linear color scale. The inset shows the distribution of the E-field in a logarithmic scale.}
\end{figure}

The same concept can be extended to conical structures, where both $R$ and $r$ are adiabatically reduced. In this case, we expect focusing as long as the gradual reduction in $r$ is greater or equal to that in $R$. Figure 4 demonstrates this effect for a cone where the groove depth $R-r=5~\mu m$ is constant, $d=50~\mu m$ and $R$ is reduced from $100~\mu m$ to $10~\mu m$ over a distance of 2 mm. As apparent from this figure, efficient deep sub-wavelength confinement to the tip ($2r=0.04\lambda$) takes place. This geometry has great promise for channeling THz radiation to micron-scale volumes for near-field imaging, spectroscopy and sensing applications.

In conclusion, we have shown that periodically corrugated metal wires can sustain spoof surface plasmon polaritons, and that their dispersion and mode profile is determined by the geometry. Apart from low-loss propagation due to the good confinement, such structures allow energy concentration and superfocusing. We expect these results to enable a new class of THz optical research and technology with only minor extensions of existing microfabrication techniques.

\begin{figure}[t]
\begin{center}
\includegraphics[width=\columnwidth]{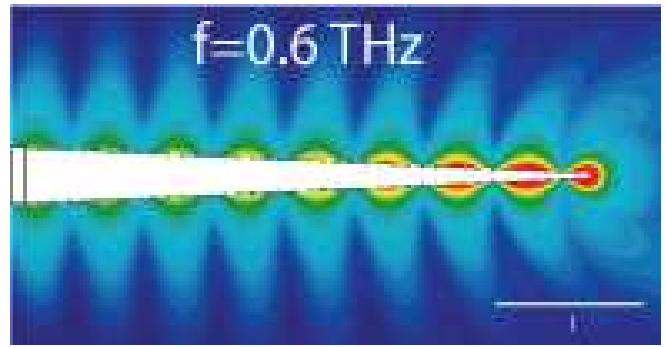}
\end{center}
\caption{Superfocusing on a corrugated cone of length $2$ mm, with constant $h=5~\mu$m and $d=50 \mu$m. $R$ is reduced from $100 \mu m$ to $10 \mu m$. Plot shows the magnitude of the E-field on a logarithmic scale spanning two orders of magnitude.}
\end{figure}

This work was supported by the Air Force Research Laboratory, under agreement number FA9550-05-1-0488, the Royal Society, the Spanish MEC under grant MAT2005-06608-C02 and the EU under project FP6-NMP4-CT-2003-505699.

\end{document}